\documentclass[twocolumn]{aastex631}

\usepackage{graphicx}
\usepackage{float} 
\usepackage{booktabs}

\begin{document}

\title{Continued Photometric Monitoring Supports Long-Term Dynamical Evolution \\ in the Young Binary Star-Disk System KH~15D}

\author{Luke Lamitina}
\affiliation{Department of Astronomy, 
California Institute of Technology, 
1200 E. California Blvd.,
Pasadena, CA 91125, USA}
\email{llamitin@caltech.edu}

\author{Lynne Hillenbrand}
\affiliation{Department of Astronomy, 
California Institute of Technology, 
1200 E. California Blvd., 
Pasadena, CA 91125, USA}
\email{lah@astro.caltech.edu}

\author{Michael Poon}
\affiliation{Department of Astronomy and Astrophysics, University of Toronto, Toronto, Ontario M5S 3H4, Canada}
\email{michaelkm.poon@mail.utoronto.ca}

\correspondingauthor{Lynne Hillenbrand}

\begin{abstract}
 We present photometric time series data spanning 2018-2024 that show the effects of temporal dynamics in the binary system KH~15D, 
 a member of the NGC 2264 star forming region. This source exhibits complex eclipsing behavior due to a precessing circumbinary disk or ring that is slightly inclined relative to the orbital plane of the binary.  Using g-band and r-band observations 
 from the Zwicky Transient Facility (ZTF) over seven observing seasons, we follow the evolution of the KH~15D lightcurve as it continues to emerge from its deepest observed photometric minimum about 15 years ago.  Our observations are consistent with previous models that propose a precessing, warped circumbinary disk orbiting KH~15D.  We verify the gradual precession of the disk by quantifying the times of eclipse ingresses and egresses. We also examine the central re-brightening within the minima of the phased lightcurve.  This feature has increased in amplitude over our observing seasons, and we measure its evolution in both amplitude and phase from year to year.  Finally, we assess color as a function of phase and brightness. Our findings support the assertion that line-of-sight variations in disk density and structure, possibly due to clumping, coupled with a precessing circumbinary disk are responsible for the central re-brightening event. 
\end{abstract}



\section{Introduction} \label{sec:intro}

KH~15D is a pre-main sequence binary star system, located in the very young open cluster NGC 2264. This system is composed of a near-equal mass binary on an eccentric orbit, with a slightly ($\sim$10 degrees) inclined
orbital plane relative to a surrounding circumbinary disk.  Observations since the original detection of the surprising lightcurve \citep{Kearns1998} have shown a long-period variation with large amplitude and a complex, evolving behavior. 

The complex eclipsing behavior of KH~15D manifests over fully half of each orbital phase,
or approximately 24 days out of the 48 day period. 
Eclipse depths exceeded 3 mag near the epoch of their discovery, but both the eclipse depth and its width/shape have evolved over time \citep{Winn2004}.

\begin{figure*}[!htb]
    \centering
    \includegraphics[width=\textwidth]{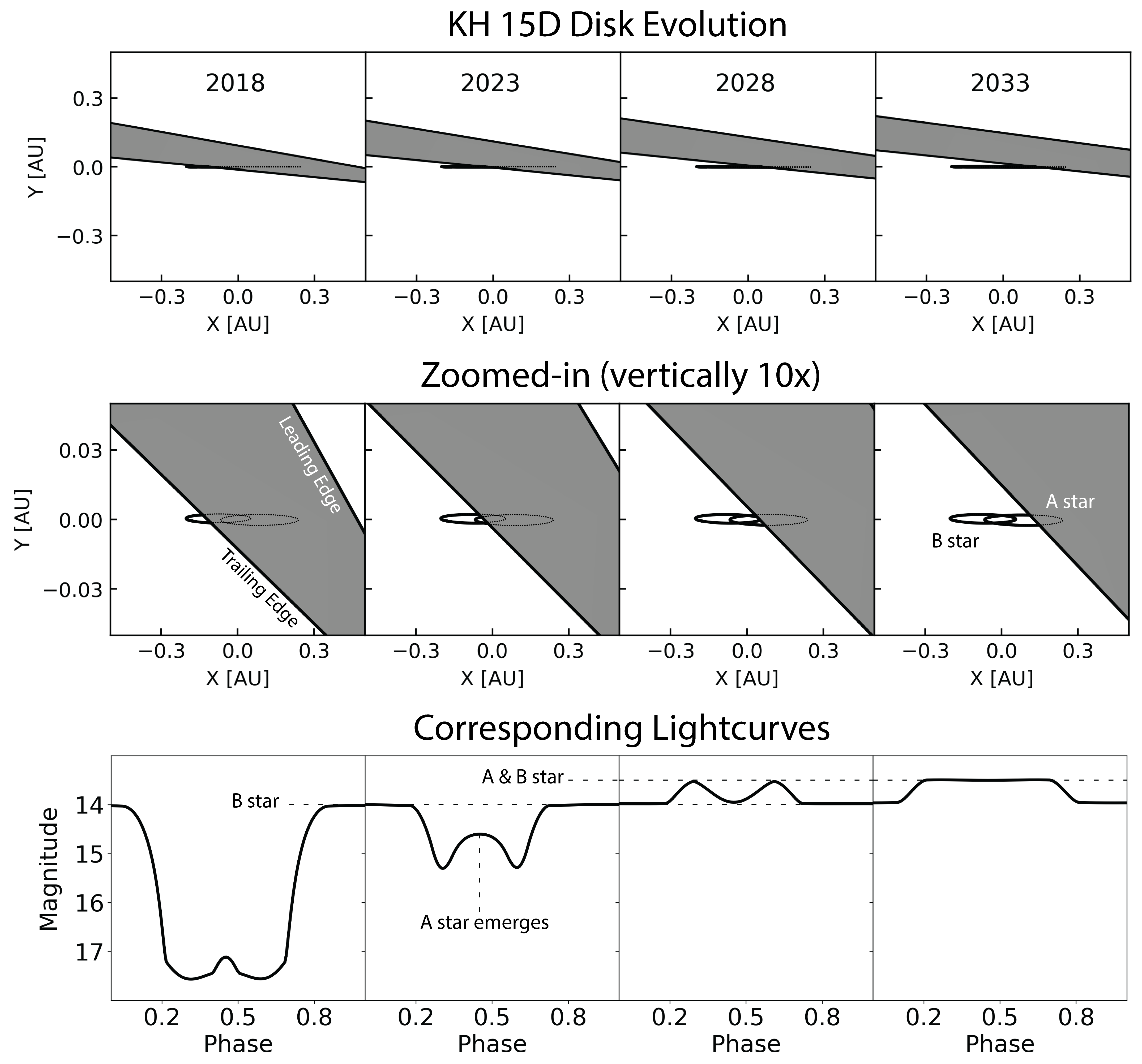}
    \caption{Illustration of the \cite{Poon2021} model for the KH-15D system.
    \textbf{Top and Middle}: To approximate the warped disk, this model uses an opaque screen bounded by a leading and trailing edge. Within each panel, the screen is nearly stationary, as the binary orbit (every 48 days) allows Star B and/or A to peek out from below the trailing edge. The dotted curves show the binary orbit behind the screen. From left to right, each edge of the screen evolves linearly in time, with the ability to rotate and move along the sky plane. This motion parameterizes the slow, rigid precession of the warped disk, where the leading (trailing) edge is the inner (outer) disk radius. Relative to the top panels, the middle panel is zoomed-in 10 times along the vertical axis. \textbf{Bottom}:  The lightcurve of KH~15D evolves substantially over a single binary orbit (within each panel) and as its warped disk precesses (across each panel). 
    A key feature in the years $\sim 2018-2023$ is the mid-phase central re-brightening, due to the emergence of Star A peeking out from behind the disk.
    }
    \label{fig:Poon2}
\end{figure*}

As demonstrated by \cite{Johnson_2004}, from 1967 to 1982 the system alternated between bright and faint states, with a phase shifted by about half a period as compared to the modern light curve. Furthermore, the lightcurve exhibited an overall brightness decrease during the 2000s, with a minimum in 2009 and 2010 during which the eclipse amplitude also decreased from nearly 4 mag to only about 2 mag \citep{herbst2010,capelo2012}. Since then, the lightcurve envelope has gradually brightened, and the eclipse amplitude has increased back to about 4 mag.   

Referencing Figure \ref{fig:Poon2}, Star A was the more visible component at the time of \cite{Kearns1998}. This source was the one peeking out above the disk, to the right of the leading edge. It is the less luminous member of the binary with an estimated mass of 0.72\(M_\odot\) \citep{Aronow2018, flaccomio1999bvri}.  Star B was the more obscured component, situated mostly behind the disk during the binary orbit. It is intrinsically the more luminous of the pair, and has an estimated mass of 0.74\(M_\odot\)\citep{Aronow2018, hamilton2001eclipses}. 
\cite{Garcia2020} postulated that, for the 2018 epochs of their data set, Star A was fully occulted by the disk, while Star B was peeking out below the disk, to the left of the trailing edge in Figure~\ref{fig:Poon2}. This more recent behavior reverses the situation described for earlier epochs.

\cite{hamilton2001eclipses} had also documented central brightening events during the eclipses, which at that time brought the system fully back to its uneclipsed brightness, near mid-eclipse. These re-brightenings were also detected in the near-infrared \citep{Kusakabe2005}. They were observed to decrease in amplitude over time \citep{Herbst2002,hamilton2005}. The long-term photometric trends in KH~15D suggest variations in the structure or density of the occulting material \citep{arul2016}.   In this paper we update the KH~15D lightcurve and describe its evolution since 2018.

Section \ref{sec:theory} outlines existing theoretical models for the geometry of the circumbinary disk and its past as well as predicted evolution over time. Section \ref{sec:observations} presents and then compares our new observational results with the \citet{Poon2021} model for KH~15D. 
We also discuss the color curve data that exhibits interesting behavior within the eclipsing events. 
Section \ref{sec:disc} contains a brief discussion and mentions other potentially similar objects. 
We present our conclusions in Section \ref{sec:conclusions}.

\section{Theoretical Models} \label{sec:theory}

Many models have been produced attempting to understand the geometry and orbital dynamics of the KH~15D binary system.
Early models {posited the presence of a precessing circumbinary disk or ring}, while later models provided {the additional} interpretation that the binary is gradually being occulted by a rigid precessing warped circumbinary disk. 

\subsection{Overview}

Theoretical geometries of the circumbinary disk included initial proposals that the system orientation was edge-on \citep{Herbst2002}. \cite{hamilton2001eclipses} suggested that the eclipses were caused by a dust feature in the circumstellar disk, orbiting the star at a semi-major axis of about 0.2 AU. This model accounted for the increasing eclipse duration and the periodic brightening events observed near mid-eclipse, which diminished over time.

\cite{Winn2004} proposed a more comprehensive model, suggesting that KH~15D is an eccentric pre-main-sequence binary system being gradually occulted by an opaque screen, likely the edge of a precessing circumbinary disk \citep{Winn2004, winn2006}. This model explained the historical lightcurve, the periodicity, the depth and duration of the eclipses, and the observed re-brightening events. 

Additional efforts to model the dynamics of the KH~15D system \citep{Arul2017,Garcia2020} have been mainly derived from the original concepts of \cite{Winn2004}. Notably, \cite{Garcia2020} give evidence that the trailing edge of the circumbinary disk seen in Figure~\ref{fig:Poon2} is actually composed of stellar sized semi-transparent clumps of dust which may explain variations in color of the lightcurve.

{The first physical theory for a warped geometry for the circumbinary disk was introduced by \cite{Chiang2004}}. Building upon work done by \cite{aly2015} and \cite{martin2017}, it was further modeled by \cite{ceppi2023}, who showed that in eccentric binaries surrounded by a disk, an additional stable configuration forms due to the alignment of the angular momentum vector of the disk to the eccentricity vector of the binary causing a polarly oriented system. 
However, tilt oscillations in the KH~15D system were studied by \citet{smallwood2019}, concluding that the system inclination was at a local minimum and will increase significantly before eventually aligning in a coplanar rather than polarly oriented configuration. 

\begin{figure*}[!htb]
    \centering
    \includegraphics[width=\linewidth]{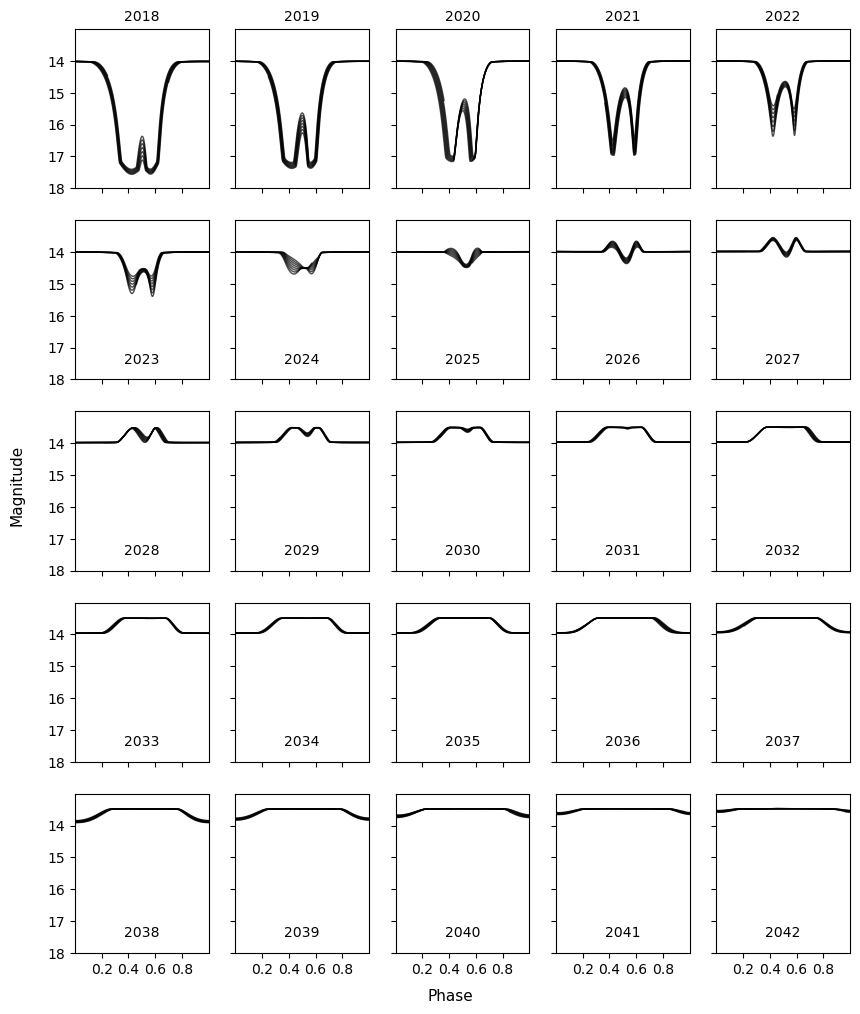}

    \caption{{Phased lightcurves produced by the knife-edge \cite{Poon2021} model extended to 2042, presented in calendar years.}
    After 2042 the lightcurve {is predicted to remain relatively flat} giving a testable ending to the KH~15D story.}
    \label{fig:Poon1}
\end{figure*}

A more detailed model for the geometry of the circumbinary disk around KH~15D was proposed by \cite{Poon2021}, motivated by a joint fit of radial velocity observations and photometry from $1955-2018$. This model aims to set certain constraints for the disk precession as well as {builds upon} the idea that the disk has a warped geometry. 

The \cite{Poon2021} model geometry features an edge-on binary orbit surrounded by a warped, circumbinary disk. This disk is misaligned relative to the binary by $\sim 5-16^\circ$, with the inner edge more inclined than the outer edge. Over the course of one binary orbital period, this system can {fluctuate} by $\sim 4$ mag, as either one or both of the binary components pass behind its optically thick disk.
As illustrated in Figure~\ref{fig:Poon2}, the \cite{Poon2021} model also predicts that after $\sim 2021$ 
($t_7$ in their model nomenclature), 
Star A also starts to peek out from below the disk, for at least some of its orbit,
in addition to Star B as previously discussed by \cite{Garcia2020}.

\subsection{Extension of the \cite{Poon2021} Model}

Yearly lightcurve predictions for KH~15D are provided in \cite{Poon2021} only until 2021. To compare with newly available photometry of KH~15D through 2024, we extend these predictions {maintaining the original parameters of the \cite{Poon2021} model.}

The schematic diagram of the KH~15D system in Figure \ref{fig:Poon2} displays the disk geometry for the years $2018-2033$ 
from the perspective of an observer, along with the corresponding schematic lightcurves. Here, the X-Y sky plane is oriented with the X-axis along the line of nodes. The \cite{Poon2021} model parametrizes the complex behavior of a warped, precessing, circumbinary disk, as an opaque screen bounded by two edges that can rotate and translate linearly in time.  

Figure \ref{fig:Poon1} shows the model lightcurve predictions up until the year 2043. 
As demonstrated by the comparison to the data (described below), this model accurately predicts several salient aspects of the modern lightcurve.
These include the gradual emergence of Star A from below the trailing edge, seen as mid-phase re-brightening events during 2018-2022, which become broader over time, 
and the gradual decrease in depth of the eclipse during 2022-2024. 

Beyond $\sim 2029$, the orbit of star B should be fully revealed, and therefore the variability over its orbit should cease. Beyond $\sim 2041$, the orbit of star A should be fully revealed, and the KH~15D system should cease to display photometric variations due to the circumbinary disk.

We now describe the data before returning to a detailed description of the model similarities and differences with the observed lightcurve.


\section{New Photometric Data from ZTF} \label{sec:observations}

Recent observations using data from the Zwicky Transient Facility \citep{Bellm2019} provide new insights into the evolution of the KH~15D system. These new measurements support the prediction that the warped circumbinary disk is gradually precessing, as modeled by \cite{Poon2021}, causing a decrease in the magnitude of the lightcurve's re-brightening events and an increase in these events' peak magnitudes. They suggest variations of structure and density within the trailing edge of the circumbinary disk contributing to the central re-brightening event as described by \cite{Garcia2020}.

\subsection{Data Acquisition and Initial Analysis}
\begin{figure}[!htb]
    \centering
    \includegraphics[width=\columnwidth]{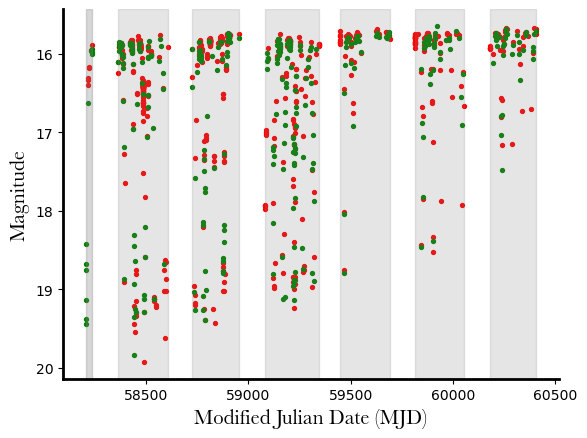}
    \caption{Combined r-band and g-band ZTF photometry of the KH~15D binary star system over the seven seasons (highlighted gray) spanning 2018-2024. An offset of +1.01 was applied to the r-band data in order to combine it with the g-band data for the following analysis.}
    \label{fig:rseasons}
\end{figure}

The photometry newly presented here consists of g-band and r-band observations that span 7 cycles or seasons between 2018 and 2024 (Figure \ref{fig:rseasons}). Each season of data contains independent observations of the source taken over a period averaging 208 days. The first season had observations spanning only 32 days, and thus it was removed from our analysis. 

The photometric measurements were retrieved from the ZTF archive \citep{Masci2019} through the IPAC lightcurve service\footnote{\url{https://irsa.ipac.caltech.edu/cgi-bin/Gator/nph-scan?submit=Select&projshort=ZTF}} in ASCII format.  The data was then filtered to retain only those photometric points with the $catflag$ value set to less than 32768. {Catalog flags are served as part of the ZTF IPAC lightcurve service and used as a metric to assess the quality of a data point. The threshold 32768 is generally defined to be the point where photometric measurements are unusable, such as due to clouds or moonlight\footnote{\url{https://irsa.ipac.caltech.edu/data/ZTF/docs/ztf_explanatory_supplement.pdf}}.} Typical errors reported on the photometry range from 0.015 mag at the bright end to 0.04 mag (r) and 0.07 mag (g) at the faint end. {The average sampling separation between observations was $\sim$ 2 days (r) and $\sim$ 6 days (g). The relative sampling between r-band and g-band observations had an average separation of $\sim$ 12 hours, meaning that when g-band data were taken, r-band data was typically obtained also within the same night or on an adjacent night.}

The ZTF data on KH~15D show the expected large amplitude brightness variations and periodicity of the lightcurve. Additionally, we call attention to the decreasing maximum depth of the eclipses in the lightcurve over the 7 seasons of data. {We also note the small increase in the maximum brightness of the lightcurve over the time span of our observations, something that the \cite{Poon2021} model predicts but not for another fifteen years (Figure \ref{fig:Poon1}).}


To increase the available amount of data in our analysis, the g-band data and r-band data were combined. In order to align their magnitudes, the average magnitude of each band during its uneclipsed sections of the lightcurve was calculated. The average magnitudes were found to be 14.92 (r) and 15.93 (g). Thus, an offset of +1.01 was applied to the r-band data (Figure \ref{fig:rseasons}). 

\begin{figure}[!htb]
    \centering
    \includegraphics[width=\columnwidth]{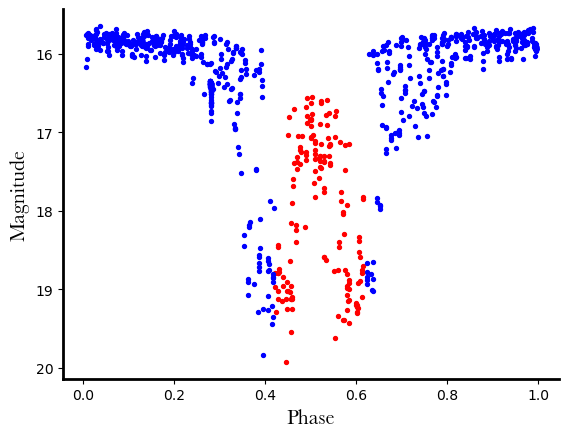}
    \caption{Phased ZTF lightcurve of KH~15D using the derived period of 48.36 days.
    {Colors} highlight the points used for fits of the central inflection (red) and points used for the fit of the broad eclipse (blue).
    }
    \label{fig:phasedr}
\end{figure}

Periodograms were produced using the standard Lomb-Scargle method to find the system's period. The period was found to be 48.36 days {which} is in agreement with the literature values.  
Using the ZTF-based period, all seasons of ZTF data were phased and the photometry was overlayed to produce a composite phased lightcurve of the system (Figure \ref{fig:phasedr}).

\begin{figure*}[!htb]
    \centering
    \includegraphics[width=\linewidth]{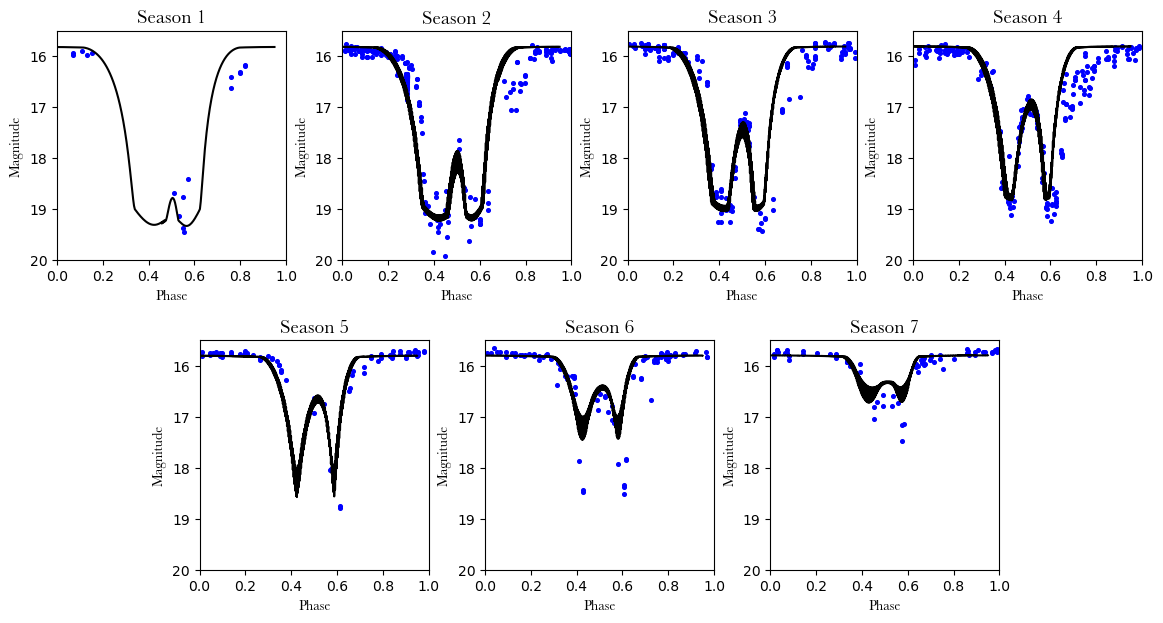}
    \caption{Phased lightcurves {with \cite{Poon2021} model (black)} of KH~15D split into seven seasons ($\sim$208 days/season) between 2018-2024, as defined in Figure~\ref{fig:rseasons}. {The model has been translated by $+1.8$ mag.} } 
    \label{fig:sevenrseasons}
\end{figure*}

In the following analysis, we examine the two main features of the KH~15D lightcurve:
the ingress/egress of the broad eclipse, and the central re-brightening that occurrs around mid-eclipse.
We separate the phase range over which we study each of the two phenomena, as illustrated in Figure \ref{fig:phasedr}. 
The ZTF data was then examined season-by-season, as represented in Figure \ref{fig:sevenrseasons}. 

\subsection{Eclipse Ingress and Egress}

As noted in previous work, while the lightcurve period has remained the same 
the eclipse duration has changed over time.  In the early 2000s, it became increasingly longer from season to season \citep{Herbst2002}. Our data taken approximately 20 years later show that the eclipse duration has been decreasing, i.e. becoming increasingly shorter from season to season. 

\begin{figure}[!htb]
    \centering
    \includegraphics[width=\columnwidth]{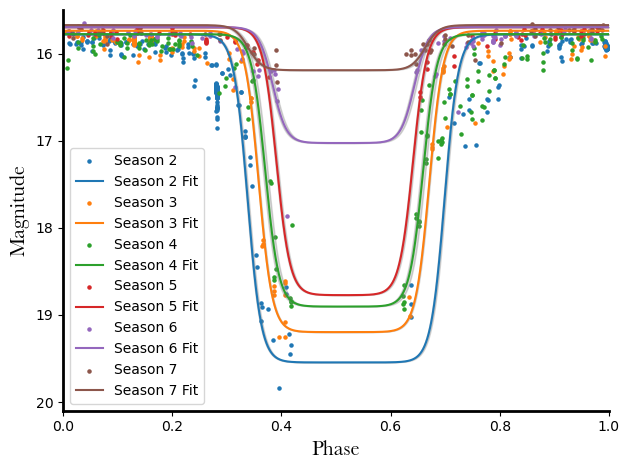}
    \caption{{Two-sided} hyperbolic tangent fits for the KH~15D data intended to quantify the ingress and egress portions of the overall lightcurve. 
}
    \label{fig:bigbump}
\end{figure}

To better understand this behavior, the season-by-season data were analyzed after removing the phases containing the central inflection ({red points in} Figure \ref{fig:phasedr}).
A {two-sided} hyperbolic tangent function was fit to the censored lightcurves (Figure \ref{fig:bigbump}), taking the average of the three faintest points\footnote{Only two, rather than three, points were used for Season 5, which has poor sampling at the faintest magnitudes} to establish the offset. While the hyperbolic tangent fits the box shape of the data better than a Gaussian would, we note that it is still not a perfect fit, especially at the eclipse minimum where there is little constraint on the true minimum. However, the fitted function is a useful form for constraining the ingress and egress parts of the eclipse, and it gives a good idea of the evolution in the absolute depth. 

From the fits, it is apparent that as the central inflection is decreasing in amplitude, the overall duration of the eclipses is also decreasing. While this general behavior can be captured by the \cite{Poon2021} model in the lightcurves of Figure \ref{fig:Poon1}, {we call attention to the disagreement in the egress shown in Figure \ref{fig:sevenrseasons}. The fact that the egress doesn't occur as rapidly as predicted by the knife-edge model is something that was initially noted by \cite{Garcia2020} and we confirm in the present data. However, the reason for why it occurs solely in the egress and not during the ingress provides another argument for clumping in the trailing edge. The relative latitude at which Star A emerges from the disk is slightly different from that where it reverses in its orbit and re-enters the disk. In a perfectly uniform knife edge disk, we would expect symmetry in the ingress and egress. However, local clumping could explain the observed scatter in the egress. It could be the case that Star A is emerging from a region with less clumping (knife-edge), but then upon re-entering the disk at a different latitude, is encountering clumping.} 

{Another possibility is that instead of the entire trailing edge being `clumpy', it could be just one single clump that is orbiting along the trailing edge, which may explain this transient egress behavior. For a clump along the trailing edge, which \cite{Poon2021} interprets as the outer disk edge at a few AU, we can estimate the orbital period of this clump. At 2-3 AU around a 1.4 solar mass binary, the orbital period would be of order $\sim$ 2-5 years. So this transient clump should change latitudes on a timescale faster than this orbit. The clumping effect is seen in Seasons 2-4, but then disappears after, which is what one would expect for an orbiting clump. }  


\subsection{Central Re-Brightening Event within the Eclipses}

The ZTF data clearly show a central inflection or reversal of the lightcurve, that occurs each phase during the eclipse. This lightcurve feature had been {present} in literature data between 1999 and 2018, but {showed an increase in amplitude} in 2018 in both the g-band and the r-band photometry from ZTF. This behavior is essentially as predicted by \cite{Poon2021}. 

\begin{figure}[!htb]
    \centering
    \includegraphics[width=\columnwidth]{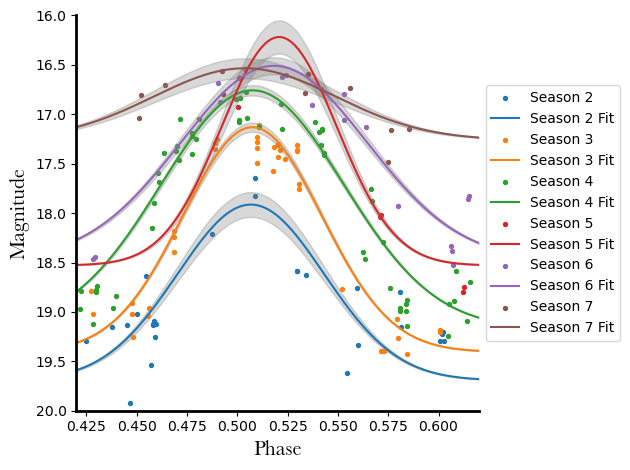}
    \caption{Central re-brightening event fit across seasons 2-7. Gray shaded region represents one standard deviation error.}
    \label{fig:bump}
\end{figure}

To better understand the nature of this re-brightening event, 
a phase constraint was applied to each season's data, isolating the inflection in the lightcurve ({red points in} Figure \ref{fig:phasedr}). Only seasons 2-7 were used due to the sparseness of the data stream in season 1 {(see Figure \ref{fig:sevenrseasons})}.  
Gaussian functions were then fitted within each season to the central re-brightening events (Figure \ref{fig:bump}) with results reported in Table~\ref{tab:inflection}. The offset, or minimum of the Gaussians, were calculated by taking the average of the 3 highest magnitude points in order to eliminate the possibility of a bad data point producing an inaccurate fit. We note that, as the seasons progress, the peak of the lightcurve inflection or central re-brightening begins to approach the uneclipsed magnitude, while the depth decreases. This observation supports the predicted lightcurves by \cite{Poon2021} (Figure \ref{fig:Poon1}).

It is also noted that the inflection is experiencing broadening. This finding from the modern KH~15D lightcurve supports the basic interpretation that the circumstellar disk is precessing in accordance with the predictions of the \cite{Poon2021} model. However, the slight difference in detail and unique time of observation (2023 is just as Star A begins to peek out from the trailing edge of the disk) {gives} us an opportunity to probe the dynamics occurring at the trailing edge of the disk.

\begin{table*}[!htb]
    \centering
    \small
    \begin{tabular}{lllllll}
        \toprule
        \textbf{Season} & \textbf{Year} & \textbf{MJD range} & 
        \textbf{Out-of-Eclipse Brightness} & \textbf{Inflection Center} & 
        \textbf{Inflection Width} & 
        \textbf{Inflection Height} \\ 
        \textbf{} & \textbf{} & \textbf{[day]} & 
        \textbf{[mag]} & \textbf{[phase]} & 
        \textbf{[phase]} &  \textbf{[mag]} \\
        \midrule
        1 &       2018 & 58204-58236 & 15.93 & \nodata & \nodata & \nodata \\ 
        2 & 2018--2019 & 58360-58607 & 15.78 & 0.50 & 0.039 & -1.72 \\ 
        3 & 2019--2020 & 58723-58956 & 15.74 & 0.51 & 0.051 & -2.34 \\ 
        4 & 2020--2021 & 59081-59345 & 15.78 & 0.51 & 0.071 & -2.72 \\ 
        5 & 2021--2022 & 59446-59693 & 15.70 & 0.52 & 0.068 & -2.99 \\ 
        6 & 2022--2023 & 59811-60052 & 15.70 & 0.52 & 0.098 & -2.92 \\ 
        7 & 2023--2024 & 60178-60407 & 15.67 & 0.50 & 0.159 & -2.94 \\ 
        \bottomrule
    \end{tabular}
    \caption{Measured quantities of the eclipsing binary star system KH~15D using newly available ZTF data. Phase values can be converted to days using a period of 48.36 days.
    }
    \label{tab:inflection}
\end{table*}

Referencing Figure \ref{fig:Poon2}, we notice that the new ZTF photometry ranging from 2018-2024 captures the predicted re-emergence of star A from the trailing edge of the disk. Since Star A is emerging by such a small amount, we are able to further investigate whether a knife-edge model is accurate for the trailing edge, or if further clumping effects as noted by \cite{Garcia2020} are active.


\begin{figure}[!htb]
    \centering
    \includegraphics[width=\columnwidth]{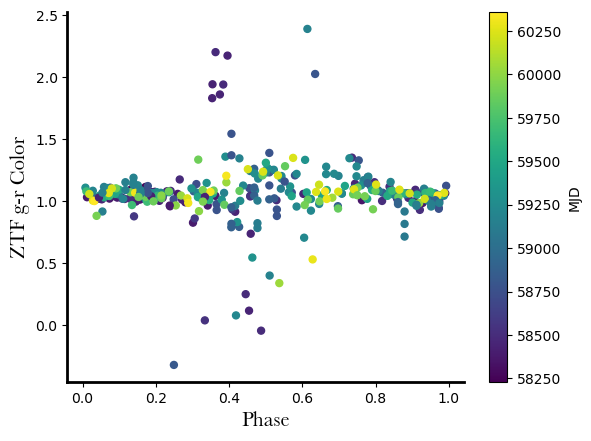}
    \includegraphics[width=\columnwidth]{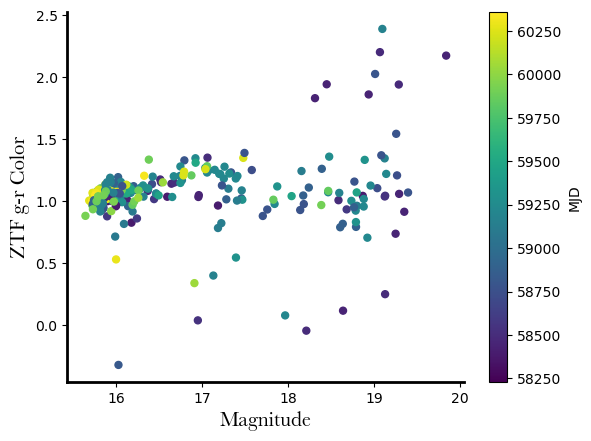}
    \caption{Color in $g-r$ as a function of phase (top) and brightness (bottom). 
    There is some amount of color variation that occurs mainly during eclipse ingress and egress, consistent with light from the stellar components of KH~15D interacting with the circumbinary material. 
    It is noted that at dimmer magnitudes this interaction is more apparent, supporting the notion of clumping within the trailing edge of the disk or ring. 
    }
    \label{fig:color}
\end{figure}

\subsection{Color Behavior}

To analytically distinguish between the knife-edge and clump possibilities, we compared the ratio of g-band to r-band fluxes in order to quantify any changes in the color of the light as a function of phase.
Any changes in color during the ingress/egress phases, or central re-brightening events, could be an indication that the light is experiencing either reddening or scattering due to dust clumping.

When $g-r$ color was plotted in relation to phase (Figure \ref{fig:color}), it was observed that the data with the largest color excursions are confined to a phase range of 0.3-0.7, which is within the dip and re-brightening phases of the lightcurve. Within this phase range it is important to note that the lightcurve is evolving rapidly in magnitude (see e.g. Figure \ref{fig:sevenrseasons}). It is therefore possible to calculate large color changes within the ingress and egress of the eclipse due to the sampling not being instantaneous. The resulting color variations would then be artifacts of the data and have no astrophysical significance.

Since the average sampling separation between r-band and g-band observations is $\sim$ 12 hours, a linear interpolation was conducted to fill in the cadence gaps between observations in the two different filters and calculate $g-r$ colors. With a period of 48 days, the ingress, egress, and re-brightening events are each  $\sim$9-12 days. Thus, there should be 
plenty of data spanning the phase range 0.3-0.7 for linear interpolation to estimate the color variation.

The color excursions in Figure~\ref{fig:color} are large and highly significant.
They can be explained if either one of the two stars is passing into or behind a region of lower density, 
causing increased scattering (bluing), and subsequently into higher density regions, where extinction (reddening) can occur. Eventually, Star B will be fully emerged over its entire orbit (see Figure~\ref{fig:Poon2}), 
and the color changes that we currently observe during the re-brightening phases will return to maintaining their non-eclipsed levels. 

It is also apparent in Figure \ref{fig:color} that this color variation tends to be more prominent in the earlier seasons. One possible reason for this is that, at these epochs, the orbit of Star A does not pass through the clumpy more transparent outer region of the disk, but passes only partially into it and then reverses out of it, along its orbit back into (or behind) the higher density regions of the disk. Essentially, since its projected orbital extreme peaks in this lower density region with more scattering, it is spending more time there. 

The large color variations seen in Figure \ref{fig:color} are in agreement in comparison with those found by \cite{Garcia2020} and suggest a model of the KH~15D system that effectively combines the knife-edge approach used by \cite{Poon2021} with a clumping aspect on the trailing edge described by \cite{Garcia2020}, 
having stellar-sized semi-transparent clumps.  We postulate that such clumps could change latitudes in the disk on a timescale that is consistent with the evolution and disappearance of the observed discrepancy between the knife-edge model and the egress portion of the lightcurves in Figure~\ref{fig:sevenrseasons}.

\section{Discussion} \label{sec:disc}

Our analysis of the recent Zwicky Transient Facility (ZTF) observations of KH~15D provides compelling evidence supporting the evolution described by both \cite{Poon2021} and \cite{Garcia2020}. The temporal evolution observed in the lightcurve—specifically, the decreasing eclipse duration and overall amplitude, and increasing duration of the central re-brightening events—aligns with the predictions of a precessing disk gradually occulting the binary system. 

The observed color changes during ingress/egress and the inflection event give insight into the fine structure within the trailing edge of the circumbinary disk. 
Understanding these density variations may provide crucial insights for refining models of disk morphology and dynamics in young stellar systems.

Future high-resolution spectroscopic and photometric observations are essential to differentiate the contributions of disk density fluctuations and stellar motions to the observed lightcurve features. Such studies will deepen our understanding of disk-star interactions and the evolution of young binary systems surrounded by protoplanetary material.

While KH~15D exhibits a large-amplitude, long-period lightcurve that is understood as being due to complex circumbinary disk dynamics, it has been noted that there are actually many close binary systems with similar behavior. KH~15D can be considered in the context of objects such as WL 4 \citep{Plavchan2008} and YLW 16A \citep{Plavchan2013}, both young stellar objects in Ophiuchus that are similar to KH~15D, and Bernhard 1 and Bernhard 2 \citep{Zhu2022}, which have unknown ages. 
All are proposed as being composed of a circumbinary disk that is tilted relative to a highly eccentric binary star system. 
Each presents a ``square wave" lightcurve with rapid descent into and emergence from a deep eclipse.  This is in contrast to
the box-like eclipses of edge-on binaries, and the sinusoidal lightcurves produced by rotation of surface starspots.
Other systems with eclipsing lightcurves that also evolve over time have only one visible star, 
with an inferred much fainter companion with a disk that periodically obscures the other.  Examples of this phenomenon 
include the famous $\epsilon$ Aur and EE Cep, and the more recently identified Bernhard 3 \citep{bernhard2024}. 

\section{Conclusions} \label{sec:conclusions}

Our analysis of recent observations from the Zwicky Transient Facility  of the intriguing young stellar object KH~15D provides strong support for the model proposed by \cite{Poon2021}, which describes the unusual eclipsing behavior as resulting from a precessing, warped circumbinary disk. The observed temporal shifts in the ingress phase of the lightcurve, along with the increasing amplitude and phase shift of the central re-brightening events, align with the \cite{Poon2021} predictions of the disk's geometry and precession rate. The re-brightening events are likely caused by the star passing through regions of varying density within the trailing edge of the disk as proposed by \cite{Garcia2020}, rather than partially emerging from the bottom of a knife-edge disk. 

This study reinforces appreciation for the complex interactions between young stellar object binary star systems and their surrounding disks, and highlights the role of disk morphology and dynamics in cases of large-amplitude periodic eclipse-like variations.
Understanding the dynamics of the KH~15D system through continued multi-color lightcurve monitoring
provides an opportunity to apply new theories towards constraining these other similar systems.
KH~15D is thus far unique, however, in the exhibition of the inflection or re-brightening that occurs roughly mid-eclipse.


\vspace{5mm}
\facility{P48(ZTF)}

\software{astropy \citep{2013A&A...558A..33A,2018AJ....156..123A},  
          }




\bibliography{bibliography}{}
\bibliographystyle{aasjournal}



\end{document}